\documentclass[amsfont,graphicx,nofootinbib,preprint,showpacs,preprintnumbers,amsmath,amssymb]{revtex4}
\usepackage{color,graphicx,epsfig}
\usepackage{ifpdf}
\usepackage{amsmath}
\usepackage{bm}
\usepackage{color}
\usepackage[english]{babel}
\usepackage{graphicx}%
\usepackage{amsfonts}%
\usepackage{amssymb}
\usepackage{braket}
\usepackage{hyperref}
\usepackage{enumerate}
\usepackage{comment}

\definecolor{nicered}{rgb}{0.7,0.1,0.1}
\definecolor{nicegreen}{rgb}{0.1,0.5,0.1}
\hypersetup{colorlinks,citecolor= nicegreen,linkcolor= nicered}

\usepackage{graphicx}
\usepackage{dcolumn}
\usepackage{bm}
\usepackage{slashed}

\begin{document}

\title{Heavy Bino and Slepton for Muon $g-2$ Anomaly}

\author{Yuchao Gu}
\affiliation{Department of Physics and Institute of Theoretical Physics, Nanjing Normal University, Nanjing, 210023, China}

\author{Ning Liu}
\affiliation{Department of Physics and Institute of Theoretical Physics, Nanjing Normal University, Nanjing, 210023, China}

\author{Liangliang Su}
\affiliation{Department of Physics and Institute of Theoretical Physics, Nanjing Normal University, Nanjing, 210023, China}

\author{Daohan Wang}
\affiliation{CAS Key Laboratory of Theoretical Physics, Institute of Theoretical Physics, Chinese Academy of Sciences, Beijing 100190, China}
\affiliation{School of Physical Sciences, University of Chinese Academy of Sciences, Beijing 100049, China}



\date{\today}

\begin{abstract}
In light of very recent E989 experimental result, we investigate the possibility that heavy sparticles explain the muon $g-2$ anomaly. We focus on the bino-smuon loop in an effective SUSY scenario, where a light gravitino plays the role of dark matter and other sparticles are heavy. Due to the enhancement of left-right mixing of smuons by heavy higgsinos, the contribution of bino-smuon loop can sizably increase the prediction of muon $g-2$ to the experimental value. Under collider and vacuum stability constraints, we find that TeV scale bino and smuon can still account for the new muon $g-2$ anomaly. The implications for LHC phenomenology are also discussed.
\end{abstract}
\maketitle


\section{Introduction}

As is known, the Standard Model (SM) is considered as a successful description of particle physics. However, many questions, such as naturalness problem and the nature of dark matter (DM) could not be solved in the SM, which call for new physics beyond the SM. The supersymmetry (SUSY) is one of the most promising extensions of the SM to solve these fundamental problems.

In addition, there also exists a long-standing anomaly, which is the anomalous magnetic moment of the muon. The anomalous magnetic moment of muon $a_{\mu}$ predicted by the SM theory~\cite{Davier:2017zfy,Blum:2018mom,Keshavarzi:2018mgv,Davier:2019can,Aoyama:2020ynm} is given by
\begin{eqnarray}
    a_{\mu}^{\rm SM}=116591810(43) \times 10^{-11}
\end{eqnarray}
Very recently, the E989 collaboration 
has reported their latest result~\cite{PhysRevLett.126.141801},
\begin{eqnarray}
    a_{\mu}^{\rm exp}=116592040(54) \times 10^{-11}
\end{eqnarray}
Then the combined result of E821 and E989 is given by~\cite{PhysRevLett.126.141801},
\begin{eqnarray}
    a_{\mu}^{\rm exp}=116592061(41) \times 10^{-11}
\end{eqnarray}
which shows about 4.2$\sigma$ discrepancy from the SM prediction,
\begin{eqnarray}
\Delta a_{\mu}=(2.51 \pm 0.59) \times 10^{-9}.
\end{eqnarray}
Such a deviation implies the light new physics related with the lepton sector. Due to the contributions of sparticles, SUSY can explain this anomaly without conflicting with other experimental measurements~\cite{Moroi:1995yh,Wang:2021bcx,Abdughani:2021pdc,Han:2020exx,Ren:2017ymm,Han:2016gvr,Martin:2001st,Abe:2002eq,Stockinger:2006zn,Endo:2011xq,Mohanty:2013soa,Ibe:2013oha,Akula:2013ioa,Endo:2013lva,Okada:2013ija,Ajaib:2014ana,Gogoladze:2014cha,Babu:2014lwa,Ajaib:2015ika,Kowalska:2015zja,Gogoladze:2015jua,Athron:2015rva,Wang:2015kuj,Kobakhidze:2016mdx,Fukuyama:2016mqb,Cox:2018qyi,Tran:2018kxv,Abdughani:2019wai,Liu:2020ser,Chakraborti:2020vjp}. On the other hand, the observed Higgs mass and null results of LHC searching for SUSY signatures have produced strong constraints on sparticle masses. Combining all these conditions, the electroweak SUSY is strongly favored, where the sleptons and electroweakinos are lighter than squarks and gluinos.

Besides, when $R$-parity is conserved, the lightest neutral electroweakino could be a good WIMP dark matter particle. If the LSP is pure wino or higgsino, the observed DM relic density requires its mass to be heavier than 1 TeV. While a pure bino DM usually overclose the Universe~\cite{Profumo:2004at,ArkaniHamed:2006mb}. The mixture of bino with a certain fraction of wino or higgsino can easily achieve the correct DM abundance~\cite{ArkaniHamed:2006mb,Abdughani:2017dqs}. However, the current limits from various direct detections are approaching to the neutrino background and have excluded most of mixture WIMP dark matter parameter space~\cite{Wang:2020coa,Aprile:2018dbl}. Therefore, the sparticles are now expected to be at TeV scale.

In this work, we consider an effective SUSY model, in which a light gravitino plays the role of dark matter and other sparticles are heavy. Due to the superweak interaction of the gravitino DM with the SM particles, the bounds from direct detections will be safely escaped. In previous works~\cite{Kobakhidze:2016mdx,Cox:2018qyi,Abdughani:2019wai,Chakraborti:2020vjp}, the masses of electroweakinos and sleptons are required to be less than about 600 GeV to accommodate the muon $g-2$ anomaly and satisfy dark matter constraints. Different from those studies, we will explore whether heavy sparticles can also explain the muon $g-2$ anomaly. We will study the bino-smuon contribution to muon $g-2$, which can be greatly enhanced by large $\mu$ and $\tan\beta$. However, such parameters may also lead to a sizable mixing in the smuon sector and jeopardize the electroweak vacuum stability~\cite{Hisano:2010re,Kitahara:2013lfa,Endo:2013lva}. We will include experimental and theoretical constraints and study the potential of solving muon $g-2$ anomaly in our scenario. The relevant collider phenomenology will be discussed as well.

\section{Benchmark Model and Muon $g-2$}
In the MSSM, provided that a pure bino as the LSP, its relic density is usually too large to satisfy the observed DM relic density because of the low annihilation rate. In order to solve this problem, the bino DM has to coannihilate with other lighter sparticles or through resonant channels, which requires the fine tuning of sparticle masses~\cite{Abdughani:2019wss,Abdughani:2019wai}. On the other hand, there also exist alternative SUSY dark matter particles, such as gravitino and axino.

The gravitino is predicted by the gauge theory of local supersymmetry, which can be a natural DM candidate in gauge-mediated supersymmetry breaking(GMSB)~\cite{Giudice:1998bp}. The gravitino mass is obtained via the Super-Higgs mechanism~\cite{Cremmer:1982en}, which strongly depends on SUSY breaking schemes. Its mass can vary from 1 eV  to 1 GeV~\cite{Giudice:1998xp}. However, the light gravitino dark matter may cause some cosmological problems~\cite{Moroi:1993mb,Asaka:2000zh,Bolz:2000fu,Roszkowski:2004jd,Cerdeno:2005eu,Pradler:2006qh,Pradler:2006hh,Gu:2020ozv}. Fortunately, the messenger particles are always predicted by the GMSB models. The lightest messenger can have interactions with the SM particles and sparticles through messenger-matter interactions or gauge interactions~\cite{Baltz:2001rq,Fujii:2002fv,Jedamzik:2005ir}. Then, the late decay of the lightest messenger to visible sector can produce a large amount of entropy so that the light gravitino relic density can be diluted to the observed value in the present universe.

In GMSB models, bino or slepton can be the next-to-lightest supersymmetric particle (NLSP), which is determined by the number of messenger generations $n$. When $n=1$, the NLSP is mainly the lightest neutralino $\tilde{\chi}^1_0$, and when $n \geq 2$, it is the lightest slepton. As a phenomenological work, we will not specify the GMSB and study the contributions of sparticles to muon $g-2$ in an effective SUSY framework, where the bino or smuons is the NLSP and other sparticles are much heavier than them.

In SUSY, the main contributions to $\Delta a_\mu$ arise from the one-loop corrections involving the smuon, sneutrino, neutralinos and charginos. The chirality flip between incoming and outgoing external muon lines can be induced by the L-R mixing in the smuon sector or SUSY Yukawa couplings of higgsinos to muon and smuon or sneutrino.  Since winos and higgsinos are too heavy in our scenario, the lightest neutralino is very bino-like. Hence, the dominant correction comes from the bino-smuon loop contribution, which is given by
\begin{eqnarray}
    \Delta a_{\mu}^{\rm bino}= \frac{g_{Y}^{2}}{8 \pi^{2}} \frac{m_{\mu}^{2} \mu {\rm tan}\beta}{M_{1}^{3}} F_{b}\left(\frac{m_{\tilde{\mu}_{L}}}{M_{1}},\frac{m_{\tilde{\mu}_{R}}}{M_{1}}\right)
\end{eqnarray}
where $g_{Y}$ is the $U(1)$ gauge coupling constant, $m_\mu$ is the muon mass, $m_{\tilde{\mu}_{L}}$ and $m_{\tilde{\mu}_{R}}$ are the mass of left-hand and right-hand smuon respectively. $\mu$ is higgsino mass parameter and $M_1$ is the bino mass. The loop function $F_b$ are given by
\begin{eqnarray}
    F_{b}(x,y)=-\frac{1}{2} \frac{N(x^{2})-N(y^{2})}{x^2-y^2},N(x)=\frac{1}{(1-x)^3} \left[1-x^2+2x{\rm log}x\right]
\end{eqnarray}
It can be seen that if the higgsino mass parameter $\mu$ is large enough, the muon $g-2$ anomaly may be explained through the bino-smuon loop contribution, due to the large smuon L-R mixing. However, we should note that such a large mixing may result in electroweak vacuum instability. In order to prevent the electroweak vacuum decay, the soft masses of smuon sector should satisfy following fitting formulae~\cite{Endo:2013lva,Kitahara:2013lfa}
\begin{eqnarray}
    \left| m_{\tilde{\mu}_{LR}}^{2} \right| \le  \eta_{\mu} \bigg[1.01 \times 10^{2}{\rm GeV} 
    \sqrt{m_{\tilde{\mu}_{L}} m_{\tilde{\mu}_{R}}}+1.01 \times 10^{2}{\rm GeV} 
    \left(m_{\tilde{\mu}_{L}}+1.03m_{\tilde{\mu}_{R}}\right)\nonumber \\ -2.27 \times 10^{4} {\rm GeV}^2+
   \frac{2.97 \times 10^6 {\rm GeV}^{3}}{m_{\tilde{\mu}_{L}}+m_{\tilde{\mu}_{R}}} 
   -1.14 \times 10^8 {\rm GeV}^4 \left(\frac{1}{m^2_{\tilde{\mu}_{L}}}+
   \frac{0.983}{m^2_{\tilde{\mu}_{R}}}\right)\bigg]
   \label{stability}
\end{eqnarray}
with
\begin{equation}
    m^2_{\tilde{\mu}_{LR}}=-\frac{m_{\mu}}{1+\Delta_{\mu}} \mu tan\beta
\end{equation} 
where $\eta_{\mu}$ is 0.88 for muon and $\Delta_{\mu}$ is the correction arising from the muon non-holomorphic Yukawa coupling, which is proportional to $\mu$tan$\beta$.

\section{numerical results}

\begin{figure}[h]
    \centering
    \begin{minipage}{8cm}
        \includegraphics[height=6cm,width=7cm]{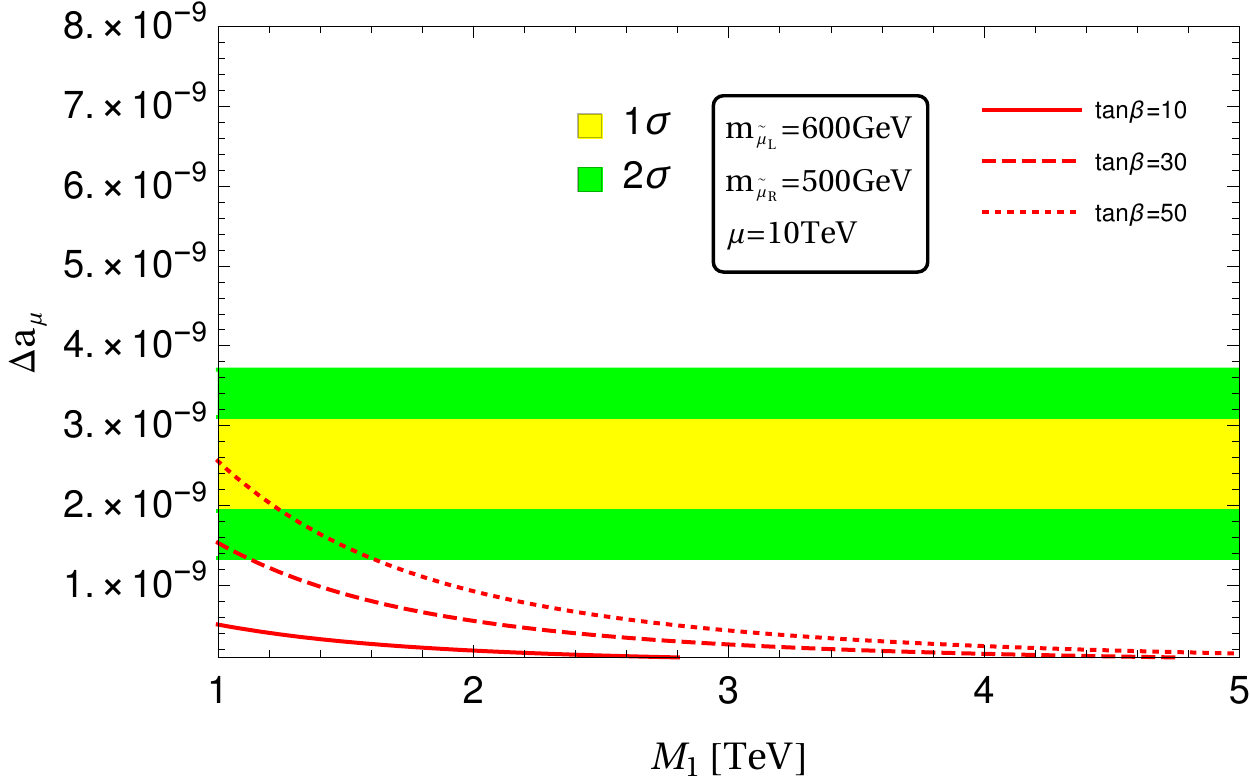}
    \end{minipage}
    \begin{minipage}{8cm}
        \includegraphics[height=6cm,width=7cm]{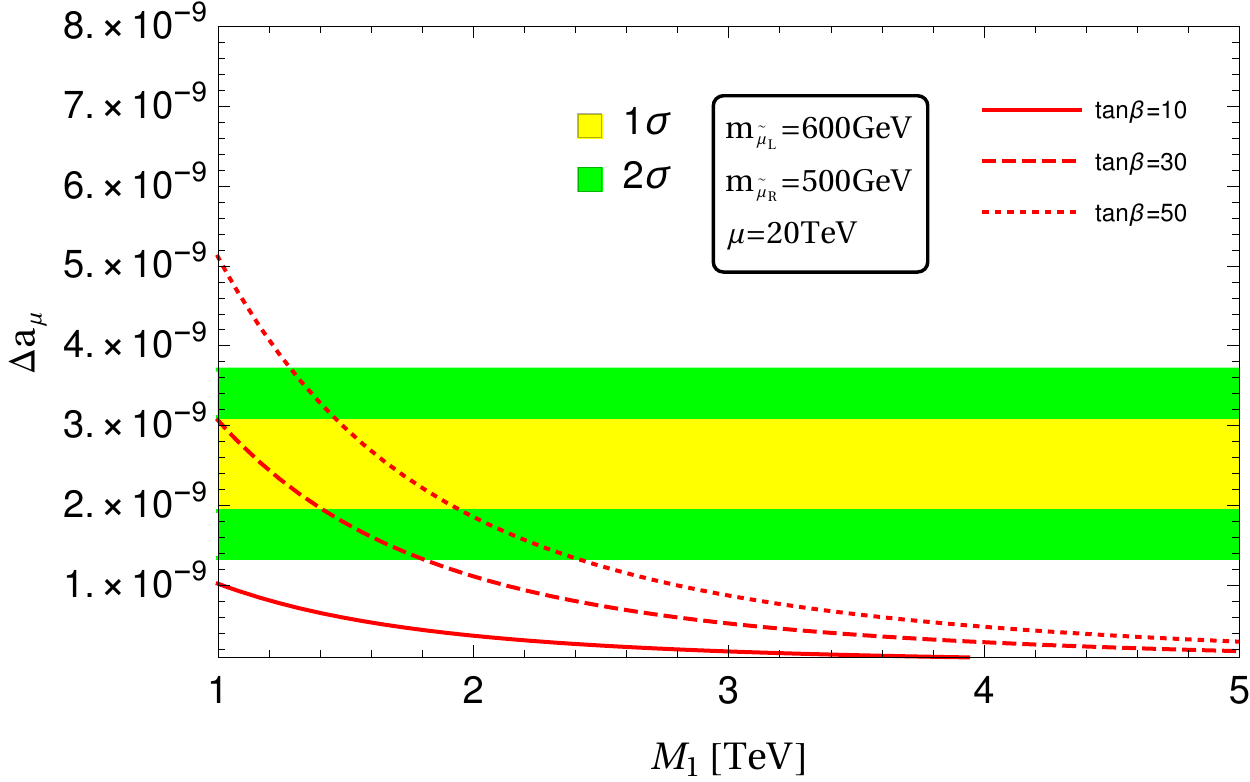}
    \end{minipage}
    \begin{minipage}{8cm}
        \includegraphics[height=6cm,width=7cm]{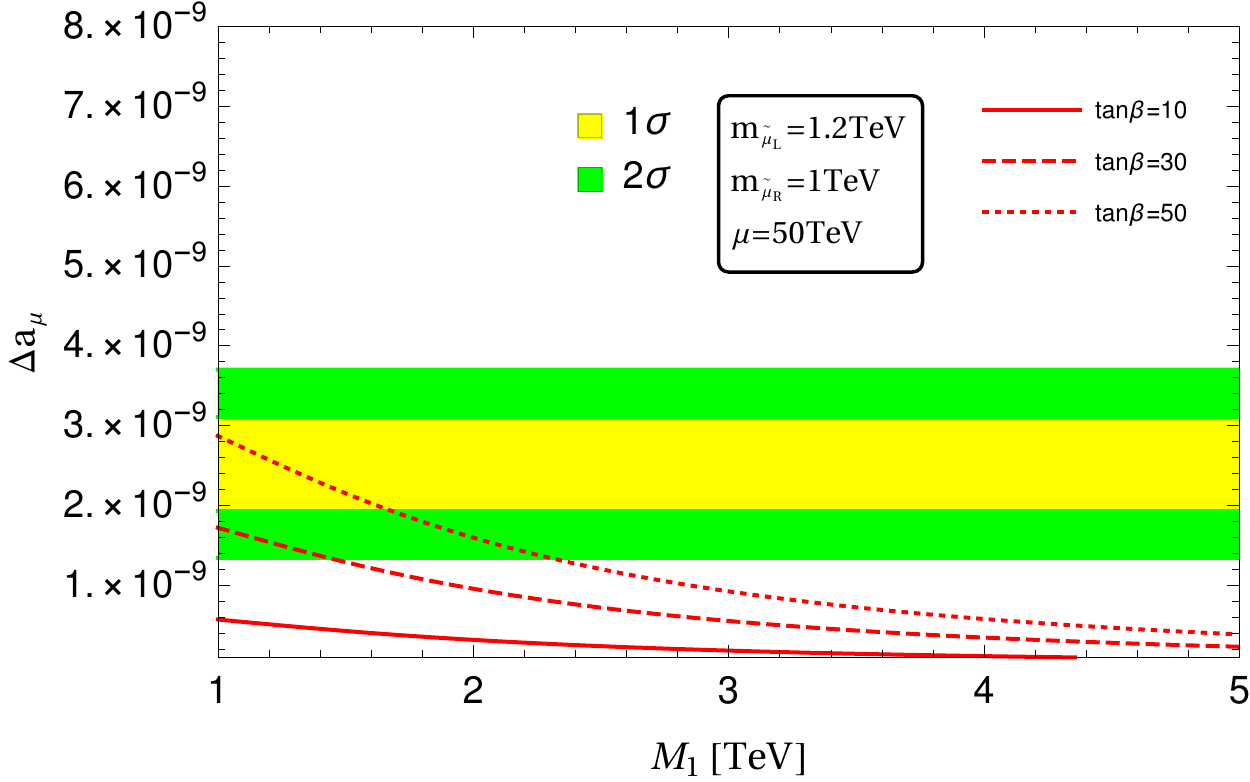}
    \end{minipage}
    \begin{minipage}{8cm}
        \includegraphics[height=6cm,width=7cm]{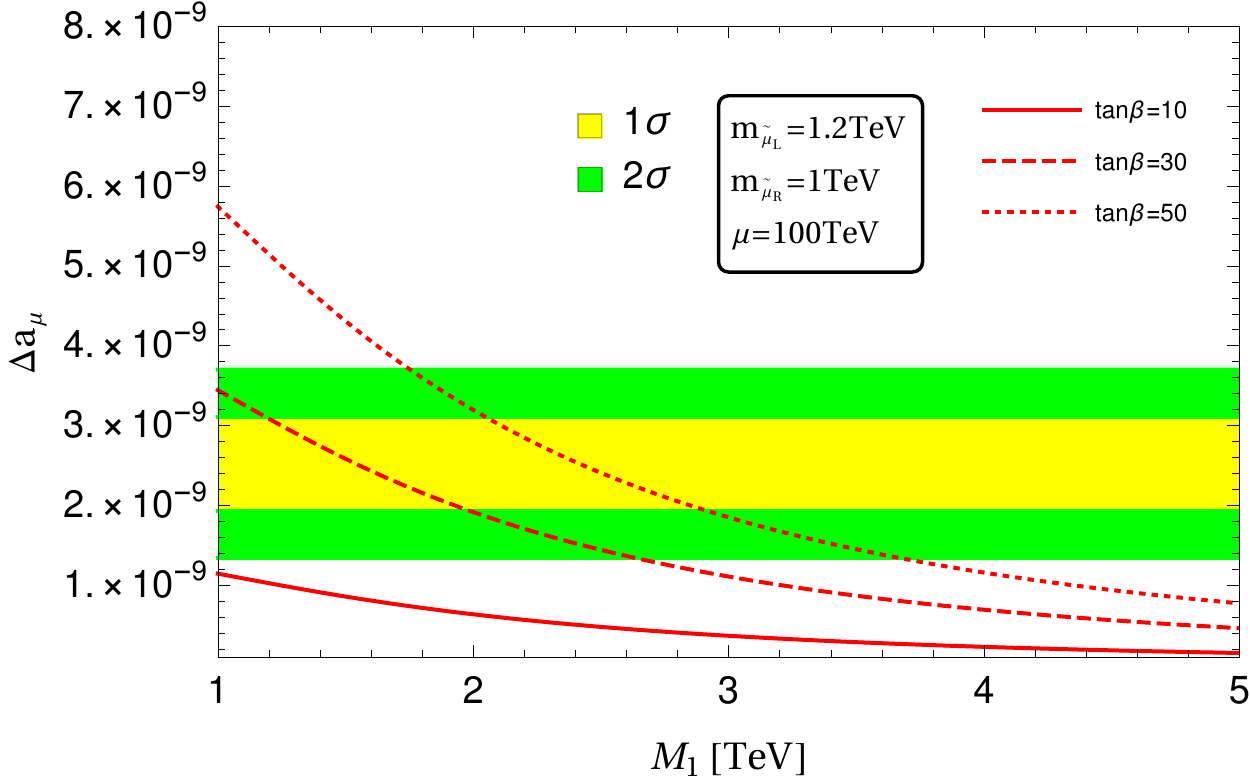}
    \end{minipage}
\caption{SUSY contribution to magnetic moment of muon $\Delta a_{\mu}$ from bino-smuon loop as the function of bino mass $M_{1}$. The yellow and green bands are $1\sigma$ and $2\sigma$ ranges for solving muon $g-2$ anomaly, respectively. Other parameters $\tan\beta=10,30,50$ and $\mu=10,20,50,100$ TeV are assumed. }
    \label{result}
\end{figure} 

Since none of colored superparticles have been observed, we set squarks and gluinos to be heavy and require them to satisfy the LHC bounds~\cite{Aad:2020aze} and Higgs mass.  Similarly, other electroweakinos (winos and higgsinos) and non-SM Higgs bosons are assumed to be heavy in this work. Therefore, in addition to light graivitino as dark matter, only binos and smuons are relatively light in our low-energy effective SUSY. The relevant model parameters are:
\begin{eqnarray}
M_1, \quad \tilde{\mu}_L, \quad \tilde{\mu}_R, \quad \mu, \quad \tan\beta
\end{eqnarray}

Besides, we note that the recent LHC search for the charginos and sleptons decaying into final states with two leptons with the missing transverse momentum has produced strong bound on the sparticle masses, which excludes the slepton mass up to 600-700 GeV~\cite{Aad:2019vnb}. However, it should be mentioned that this constraint is obtained by assuming three generations of mass-degenerate sleptons, which is not suitable for our case. When the selectron and smuon massed are separated, the lower bounds on the right-handed smuon is 450 GeV and left-handed smuon is 560 GeV (as shown in Fig.8 of Ref.~\cite{Aad:2019vnb}). Considering these bounds, we require the right-handed smuon is heavier than 500 GeV and the left-handed smuon is heavier than 600 GeV in our calculations.

The numerical results are presented in Fig.~\ref{result}, where all the parameters satisfy the electroweak vacuum stability condition in Eq.~\ref{stability} and the LHC bounds. In the upper panels, we show the results for the electroweak scale smuons, $m_{\tilde{\mu}_L}=600$ GeV, $m_{\tilde{\mu}_R}=500$ GeV that are still allowed by the current LHC searches for sleptons. When higgsino mass parameter $\mu=10$ TeV, it can be seen that the value of $\tan\beta$ has to be larger than 30 in order to satisfy the experimental value of muon $g-2$ in $2\sigma$ range. Larger $\tan\beta$ and $\mu$ can further increase SUSY correction $\Delta a_\mu$ and thus enhance the theoretical prediction to be consistent with the measurement within $1\sigma$ range. In the lower panels, we show the results for TeV scale smuons, $m_{\tilde{\mu}_L}=1.2$ TeV, $m_{\tilde{\mu}_R}=1$ TeV. Similar to the upper panels, but since the bino-slepton loop contribution is suppressed by heavy sleptons, the higgsino mass parameter $\mu$ is required to be larger to explain the muon $g-2$ anomaly for the same $\tan\beta$. For all panels, we also note that SUSY contributions to $\Delta a_\mu$ can be still sizable even the bino mass is heavier than 1 TeV as long as $\mu$ and $\tan\beta$ are large enough but without violating the vacuum stability.

Finally, we discuss the possible LHC phenomenology of heavy bino and slepton in our scenario. For slepton as NLSP, it can be produced in pair through the Drell-Yan process $pp \to \tilde{\ell}^+\tilde{\ell}^-$ at the LHC. Then, slepton will decay to gravitino and lepton, whose decay width is given by
\begin{eqnarray}
\Gamma(\tilde{\ell} \to \ell \tilde{G})=\frac{m^5_{\tilde{\ell}}}{6m^2_{3/2}M^2_{pl}}\left[1-\frac{m^2_{3/2}}{m^2_{\ell}}\right]^4.
\end{eqnarray}
Since the slepton is much heavier than gravitino, such a decay will happen inside the detector. Therefore, the final signature will be the dilepton plus missing energy events at the LHC. The HL-LHC will increase present mass reach by 20-50\%. On the other hand, if the NLSP is bino, it can be produced through the $t-$channel process $pp \to \tilde{\chi}^0_1 \tilde{\chi}^0_1$ via exchanging a squark and promptly decay to gravitino and photon, whose decay width is given by
\begin{eqnarray}
\Gamma(\tilde{\chi}^0_1 \to \gamma \tilde{G})=\frac{m^5_{\tilde{\chi}^1_0}c^2_W}{16\pi m^2_{3/2}M^2_{pl}}
\end{eqnarray}
The resulting signature will be two photons plus missing energy events at the LHC. However, due to the current strong constraint on squark mass, the prodcution rate of this process is much suppressed. Besides, there can be a two-step decay process $pp \to \tilde{\ell}^+\tilde{\ell}^- \to \ell^+\ell^-\tilde{\chi}^0_1\tilde{\chi}^0_1 \to \ell^+\ell^-\gamma\gamma+\slashed E_T$. The future higher colliders, such as HE-LHC and FCC-hh/SPPC, may be able to search for this distinctive signature.                  

\section{Conclusions}
The new combined muon $g-2$ experimental result strongly indicates the new physics beyond the SM. Supersymmetry as a leading extension of the SM provides a possible way to accommodate such an anomaly. Light electroweakinos and smuons are usually needed to enhance the prediction of $g-2$ and satisfy dark matter constraints in SUSY models. In this work, we investigate potential of heavy sparticles for contributing to muon $g-2$ anomaly and focus on the bino-smuon loop correction in an effective SUSY scenario. A sizable left-right mixing of smuons induced by large $\tan\beta$ and higgsinos mass parameter $\mu$ in bino-smuon loop can greatly enhance muon $g-2$ to reach the combined experimental value. Considering the constraints from the LHC searches for sparticle and the electroweak vacuum stability, we note that the bino and slepton with the masses at TeV scale can still account for the muon $g-2$ anomaly. We also discuss the possible LHC signatures in our scenario.

\section{acknowledgments}
We thank Lei Wu and Bin Zhu for their helpful suggestions and discussions. This work is supported by the National Natural Science Foundation of China (NNSFC) under grant Nos. 117050934 and 11805161.
\bibliography{refs}

 \end{document}